\documentclass[final]{cvpr}

\usepackage{times}
\usepackage{epsfig}
\usepackage{graphicx}
\usepackage{amsmath}
\usepackage{amssymb}
\usepackage{booktabs}
\usepackage[labelsep=period]{caption}
\usepackage{graphics}

\usepackage[pagebackref=true,breaklinks=true,colorlinks,bookmarks=false]{hyperref}

\def\cvprPaperID{****} % *** Enter the CVPR Paper ID here
\def\confYear{CVPR 2021}

\begin{document}

%%%%%%%%% TITLE
\title{CISRNet: Compressed Image Super-Resolution Network}

\author{Agus Gunawan, Sultan Rizky Hikmawan Madjid\\
{\tt\small \{agusgun, suulkyy\}@kaist.ac.kr}
}

\maketitle

\begin{abstract}
% Abstract is optional. This will not affect the final score.
In recent years, tons of research has been conducted on Single Image Super-Resolution (SISR). 
However, to the best of our knowledge, few of these studies are mainly focused on compressed images.
A problem such as complicated compression artifacts hinders the advance of this study in spite of its high practical values.
To tackle this problem, we proposed CISRNet; a network that employs a two-stage coarse-to-fine learning framework that is mainly optimized for Compressed Image Super-Resolution Problem.
Specifically, CISRNet consists of two main subnetworks; the coarse and refinement network, where recursive and residual learning is employed within these two networks respectively. 
Extensive experiments show that with a careful design choice, CISRNet performs favorably against competing Single-Image Super-Resolution methods in the Compressed Image Super-Resolution tasks.
\end{abstract}

\section{Introduction}

Deep Learning (DL) is used in the domain of computer vision for various fields such as image colorization \cite{cheng2015deep}, classification \cite{khosla2020supervised}, etc.
One field that hugely benefits from this is  Single Image Super-Resolution (SISR); a classical computer vision problem that aims to generate an accurate and visually pleasing high-resolution (HR) image given its degraded low-resolution (LR) measurement counterpart \cite{o2019deep}.
SISR has been incorporated into various real-world applications such as security \& surveillance imaging \cite{zhang2010super}, medical imaging \cite{bates2007multicolor}, etc. 
The problem is fundamentally ill-posed since, for a given low-resolution (LR) image, more than one possible high-resolution image can be generated with slight variation in camera angle, color, brightness, and other variables \cite{bashir2021comprehensive}.

To tackle this inverse problem, plenty of image super-resolution algorithms has been proposed. Among them, Residual Dense Network (RDN) \cite{zhang2018residual} and Residual Channel Attention Network (RCAN) \cite{zhang2018image} are the current state-of-the-art. 
However, most of these state-of-the-art methods have limited applicability for real-world usage such as remote sensing image processing, where most images incorporated for these purposes are in form of lossy, compressed images.

\begin{figure}[t]
\centering
\includegraphics[width=\columnwidth]{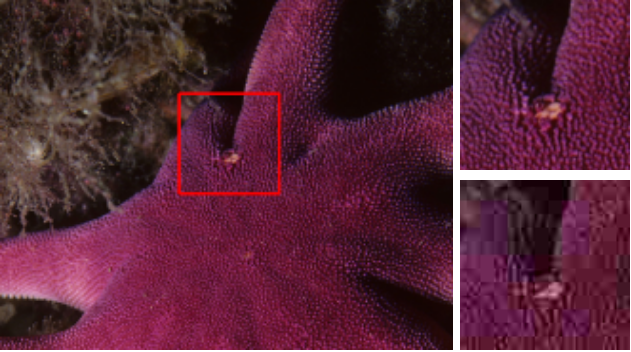}
   \caption{Comparison between a lossless and lossy image. The left reference image is lossless. The top patch is the patch extracted from the lossless image, while the bottom patch is extracted from the lossy image. Notice how the top patch looks more visually pleasant and contains fewer artifacts compared to the bottom patch.}
\label{fig:introduction_png_vs_jpeg}
\end{figure}

Figure \ref{fig:introduction_png_vs_jpeg} gives an example of a comparison between a lossless and lossy image, and we can see how the lossy image contains more noise and artifacts compared to its lossless counterpart.
Although tons of research has been conducted to tackle the SISR problem and plenty of effective SISR methods have been proposed over the past few decades, few methods were mainly concerned with compressed images \cite{xiao2012single, ono2013optimized}. 
The core issue of compressed images SR is how to reduce compression noise and preserve details as much as possible when enhancing image resolution \cite{chen2018cisrdcnn}. 

In this paper, we are trying to tackle the problem in a different direction, where instead of using a lossless image where details are preserved and image quality is maintained, a compressed image is incorporated. 
Based on the above insights, we propose a Compressed Image Super-Resolution Network (CISRNet) which is mainly optimized for compressed images. 
CISRNet employs a coarse-to-fine learning framework consists of a coarse super-resolution network and refinement network. In each network, we use a Residual-in-Residual Group structure similar to RDN \cite{zhang2018residual}. We also derive two different dual-attention blocks from Convolutional Block Attention Module (CBAM) \cite{woo2018cbam} to be used in each Residual Group (RG) of Residual-in-Residual Group (RIRG). The goal of the dual-attention block is to better emphasize meaningful features along two principal dimensions; channel \& spatial axes.

In summary, the contributions of this corresponding study are three-fold:
\begin{enumerate}
    \item First, we tackle the problem of SISR for lossy, compressed images which are more applicable for practical application in everyday life in the place of lossless, uncompressed images. 
    \item Second, we propose a new network for compressed Image Super-Resolution, where it employs a coarse-to-fine learning framework, and each network use RIRG block. 
    \item Finally, we experiment on the existing super-resolution methods and our proposed method using compressed images and empirically found that the CISRNet outperforms other SISR methods in these compressed images based on its qualitative and quantitative measurements for the more complex evaluation dataset. Meanwhile, it achieves competitive performance for a less complex evaluation dataset.
\end{enumerate}

\section{Related Work}

\noindent \textbf{Single Image Super-Resolution.} Deep learning-based SISR methods have been widely proposed and achieve state-of-the-art performance in terms of either peak signal-to-noise ratio (PSNR) or visual quality.
% Single Image Super-resolution (SISR) refers to the problem of approximating a high resolution ($I_{HR}$) image from its corresponding low resolution ($I_{LR}$) input image. 
A milestone work that introduced CNN into SR was proposed by Dong \textit{et al.} \cite{dong2014learning}, where a three-layer fully convolutional network was trained to minimize the mean squared error between the SR image and the original HR image. 
Soon after, the network architecture is constantly improving. 
By introducing residual learning to ease the training difficulty, Kim \textit{et al}. proposed deeper models VDSR \cite{kim2016accurate} and DRCN \cite{kim2016deeply} with more than 16 layers. 
To learn high-level features without introducing an overwhelming number of parameters, recursive learning is also introduced in the SR field \cite{wang2020deep}. 
Moving forward, for the sake of fusing low-level and high-level features to provide richer information and details of reconstruction, RDN \cite{zhang2018residual}, and ESRGAN \cite{wang2018esrgan} also adopted dense connections in layer-level and block-level.

Furthermore, Zhang \textit{et al.} proposed the Residual Channel Attention Network (RCAN) \cite{zhang2018image} by introducing the channel attention mechanism into a modified residual block for advanced image SR to learn features across channels and enhance long-term information by exploiting abundant low-frequency information for super-resolution. 
Notice how these networks only perform on typical degradation (e.g. bicubic, bi-linear) and forgo or naively train their models on other distortions, which frequently appear in daily multimedia devices, such as noises and JPEG compression.
This results in sub-par performance for the existing image super-resolution networks on unseen distortions.

\noindent \textbf{Compression Artifact Removal.} Several approaches on compression artifact removal have been addressed for several years \cite{liu2020comprehensive}. 
The vast majority of approaches can be classified as processing-based \cite{foi2007pointwise, wong2009document}, which typically rely on information in the Discrete Cosine Transform (DCT) domain, and others are learning-based \cite{dong2015compression, svoboda2016compression}.

Following the success of the application of Deep Convolutional Neural Networks (Deep-CNNs) in image processing tasks, such as image denoising \cite{zhang2017beyond} and Single-Image Super-Resolution \cite{dong2015image}, Deep-CNNs have been applied with success to JPEG compression artifact removal task. 
In this case, the objective is to map degraded images into distribution without the presence of noise. 
The first attempt with this kind of model has been done by Dong \textit{et al.} \cite{dong2015compression} who proposed the ARCNN, a model inspired by SRCNN \cite{dong2015image}, a neural network for Super-Resolution. 
% Later, Wang \textit{et al.} proposed D3 \cite{wang2016d3}, a deep neural network that adopts JPEG-related priors to improve construction quality which obtained an improvement in speed and performances compared to the previous models.

Recently, \cite{zhang2018dmcnn} proposed DMCNN, a Dual-Domain Multi-Scale CNN, which gains higher results quality than the previous works, by using both pixel and frequency (\textit{i.e.} DCT) domain information.
Even though recent advances show how deep-learning-based methods for image compression artifact removal, it is hardly incorporated hand-to-hand with the Single-Image Super-Resolution method that matches with our objective in this paper. 
Due to this, our proposed network tries to tackle both of these problems simultaneously.

\noindent \textbf{Coarse-to-Fine Framework.} Coarse-to-fine processing is an integral part of efficient algorithms in computer vision. 
Iterative image registration \cite{lucas1981iterative} gradually refines registration from coarser variants of the original images, while in \cite{hu2016efficient} a coarse-to-fine optical flow estimation method is proposed. 
Efficient action recognition is achieved in \cite{wu2019liteeval} by using coarse and fine features coming from two LSTM (Long Short-Term Memory) modules. 
In \cite{ma2019coarse} coarse-to-fine kernel networks are proposed, where a cascade of kernel networks are used with increasing complexity. 
Existing coarse-to-fine methods consider both coarse input resolution, as well as gradually refined processing. 
Recently, Tian \textit{et al.} \cite{tian2020coarse} proposed a coarse-to-fine super-resolution CNN (CFSRCNN) for single-image super-resolution, where it combines low-resolution and high-resolution features by cascading several types of modular blocks to prevent possible training instability and performance degradation caused by the upsampling operation.

\section{Proposed Method}
In this study, we propose a network to solve the problem of Single Image Super-Resolution (SISR) for compressed images, where a recursive learning scheme is employed under a two-stage coarse-to-fine learning framework. We name our network CISRNet.

\begin{figure*}[t]
\centering
\includegraphics[width=\textwidth]{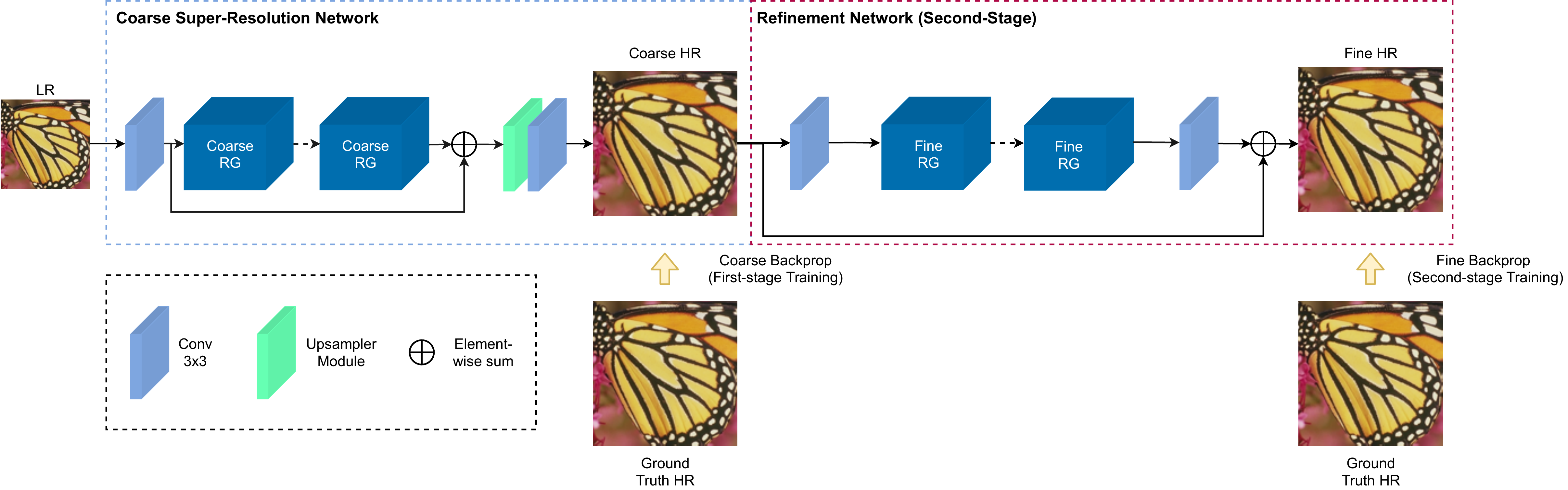}
   \caption{The network architecture of our Compressed Image Super-Resolution Net (CISRNet). The network employs a two-stage coarse-to-fine learning framework. Recursive learning schema is employed for the coarse super-resolution network, while residual learning is employed for the refinement network.}
\label{fig:cisr_net}
\end{figure*}

\subsection{Network Structure}
Figure \ref{fig:cisr_net} shows our CISRNet. 
We employ a two-stage coarse-to-fine learning framework within our network. 
The detail of the framework can be seen in Section \ref{sec:coarse_to_fine_learning_framework}. 
The network consists of two main sub-network which are coarse super-resolution network and refinement network. 
The coarse super-resolution network uses a similar approach in \cite{zhang2018image, zhang2018residual}. 
The main function of this sub-network is to create a high-resolution image and reduce the artifact of the compressed image simultaneously. 
Suppose $I_{LR}$ and $HR_{coarse}$ denotes the input and output of the coarse SR network. The input image $I_{LR}$ will go to the first part where the shallow feature $F_{shallow}$ is extracted. 
This feature is used in the next part which is a coarse residual-in-residual group consisting of several coarse residual groups (Coarse RG) blocks to extract the deep feature $F_{deep}$ of the low-resolution image. 
The detail of Coarse RG can be seen in Figure \ref{fig:coarse_rg}. 
This deep feature will be used by the upsampler module to upsample the spatial resolution of the feature map $F_{upsampled}$. 
We employ pixel-shuffle layer and convolutional layer \cite{shi2016real} in the upsampler module. 
After that, the last part of the coarse SR will reconstruct $F_{upsampled}$ to create a coarse super-resolution result $HR_{coarse}$. 
The coarse super-resolution network will result in a high-resolution image where the noisy and blocky artifacts of the compressed image are reduced.
However, the result can still have some leftover noise which may contain some cues to reconstruct high-frequency details.
This corresponding image is then used as the input of the subsequent network known as the refinement network.

The refinement network is mainly responsible to reduce the leftover noise, reconstruct high-frequency detail and texture, and sharpen the coarse high-resolution image to make it look visually appealing. 
This sub-network employs residual learning similar to the study in \cite{zhang2017beyond}. 
The network will mainly learn residual features of the image which will eventually reduce noise and sharpen the input image. 
The output $HR_{fine}$ can be formulated as.
\begin{equation}
    \label{eq:output_of_the_second_subnetwork}
    HR_{fine} = HR_{coarse} + F_{refinement}(HR_{coarse})
\end{equation}
This sub-network ($F_{refinement}$) employs a single convolutional layer to output a shallow feature representation. The shallow feature representation is used as the input of several fine residual group (Fine RG) blocks. 
The detail of the Fine RG block can be seen in Figure \ref{fig:fine_rg}. 
The output of this block is a deep feature representation that will be used in the last convolutional layer to create a residual image that can reduce the noise, enhance, and sharpened the image.

\subsection{Two-Stage Coarse-to-fine Learning Framework}
\label{sec:coarse_to_fine_learning_framework}

A two-stage coarse-to-fine learning framework has been widely used in image inpainting \cite{ma2019coarse, yu2018generative} and is effective. 
We argue that the problem of Single Image Super-Resolution (SISR) for compressed images has similar nature with image inpainting. 
This is because compressed images have many blocky artifacts which contain a homogeneous and unnatural texture. 
We can think of this homogeneous texture as a mask because it does not have any important information except the color. 
It can also contain unnatural texture which represents the texture incorrectly. 
The sample of unnatural texture can be seen in Figure \ref{fig:introduction_png_vs_jpeg} where some blocky artifacts have incorrect texture. 
Because of this similarity, we also employ a two-stage coarse-to-fine learning framework that achieves some success in image inpainting problems.

The learning framework will only train the coarse super-resolution network for the first stage. After that, the second stage of the training will train coarse super-resolution network and refinement network jointly by using the supervision from the refinement network output.
The coarse super-resolution network is trained to achieve one goal; reduce the noisy and blocky artifact of the compressed image.
To achieve this, we employ $L1$ loss for the first stage training.
This loss can maximize the peak-signal-to-noise ratio (PSNR) of the high-resolution image but has a limitation where it over-smooth the texture and lack high-frequency details \cite{ledig2017photo}.
To prevent this limitation and explicitly leave some noise that may important to reconstruct high-frequency details, we assign a low weight of 0.1 for the loss in the first stage of training.
The $L1$ loss between coarse high-resolution image and the ground-truth high-resolution image can be formulated as.

\begin{equation}
    \label{eq:l1_loss_formulation}
    L1(\hat{I}_{c}, I_{GT}) = \dfrac{1}{hwc}\sum_{i,j,k} \lvert \hat{I}_{c(i,j,k)} - I_{GT(i,j,k)} \rvert
\end{equation}

\noindent where $\hat{I}_{c}$ and $I_{GT}$ denotes the high-resolution image from coarse super-resolution network and the ground-truth high-resolution image respectively.

In the second stage of training, the training will train the full network consists of coarse super-resolution network and refinement network.
The full network is trained to generate a high-resolution image by using the coarse super-resolution network where the refinement network will eliminate left-over noise, sharpening the image, and reconstruct high-frequency detail and texture. 
This is all done jointly in one epoch.
To achieve these goals, the loss function used for this training is the content loss, which is the combination between $L1$ and perceptual loss. 
The content loss can be formulated as.

\begin{equation}
    \label{eq:content_loss_formulation}
    \resizebox{.9\hsize}{!}{%
    $L_{content}(\hat{I}_{f}, I_{GT}) = \lambda_{L1}L1(\hat{I}_{f}, I_{GT}) + \lambda_{p}L_{p}(\hat{I}_{f}, I_{GT})$%
    }
\end{equation}
\noindent where $\lambda_{L1}$ and $\lambda_{p}$ denotes the weight for $L1$ loss and perceptual loss respectively, and $\hat{I}_{f}$ denotes the result of refinement network. 
This combination of pixel-wise loss and perceptual loss was found to be effective to produce an image with rich and detailed texture \cite{ledig2017photo}. 

We use the pre-trained VGG19 \cite{simonyan2014very} layer for the perceptual loss $L_{p}$ where we use the feature maps from the deeper network layer. 
Specifically, we use the VGG feature maps $\phi_{5,4}$. 
This setting is similar to the perceptual loss setting in \cite{ledig2017photo}. 
The perceptual loss is formulated as $L2$ distance between the high-level representation of two images:

\small
\begin{equation}
    \label{eq:perceptual_loss_formulation}
    L_{p}(\hat{I}_{f}, I_{GT}) = \dfrac{1}{hwc}\sum_{i,j,k}(\phi(\hat{I}_{f})_{(i,j,k)} - \phi(I_{GT})_{(i,j,k)})^2
\end{equation}

\subsection{Residual Group (RG)}
\label{sec:residual_group}

\begin{figure}
    \centering
    \includegraphics[width=.8\columnwidth]{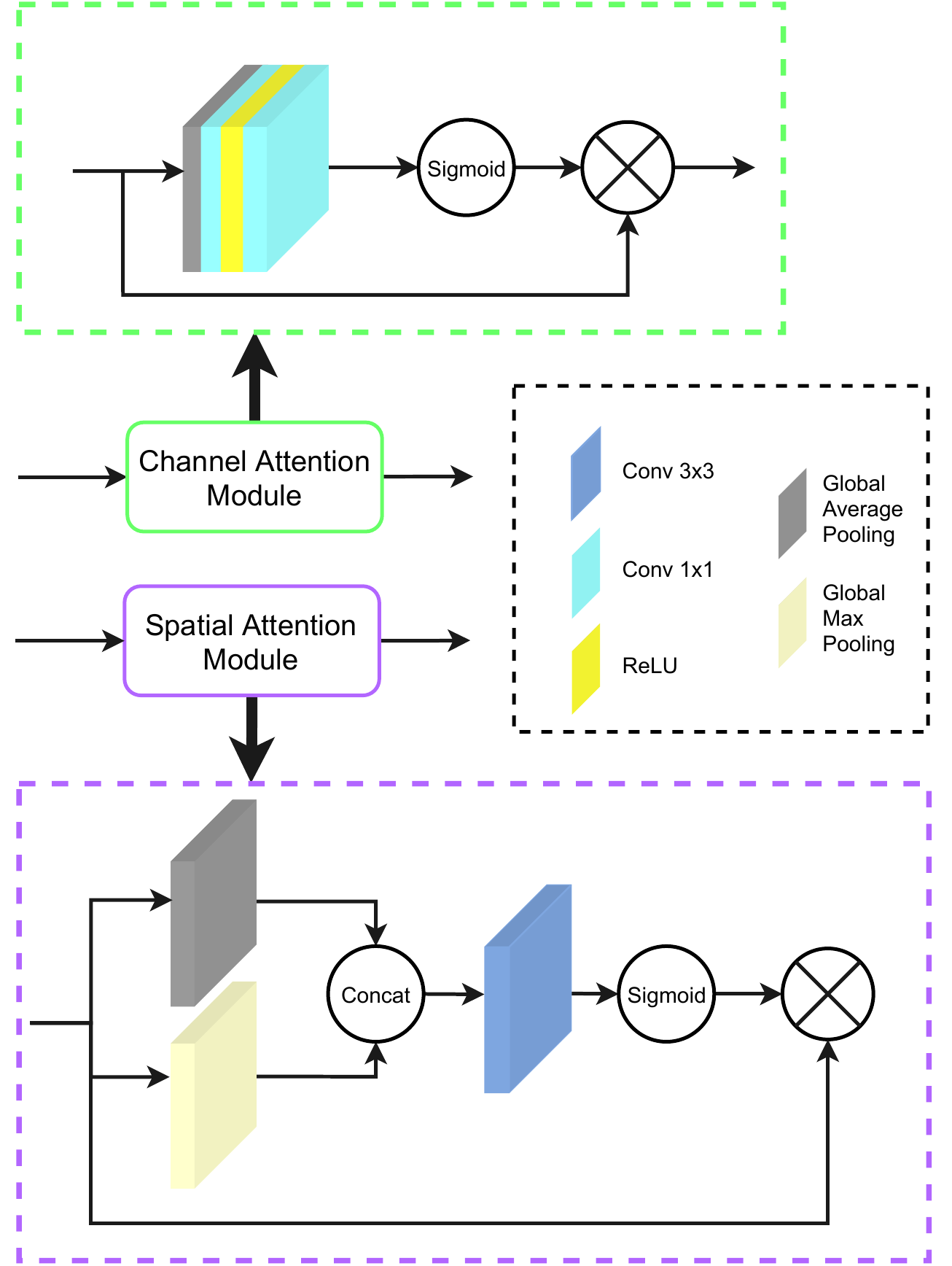}
    \caption{Channel attention and spatial attention module \cite{woo2018cbam}.}
    \label{fig:attention_module}
\end{figure}

The main basic block for the coarse super-resolution network and the refinement network is the Residual Group (RG) which uses a channel and spatial attention module. 
These two attention modules are adapted from the study by Woo \textit{et al} \cite{woo2018cbam}. 
The channel attention module will learn 'what' to attend in the feature map, while the spatial attention module will learn 'where' to attend in the feature map. 
The architecture of these two modules can be seen in Figure \ref{fig:attention_module}. 

\begin{figure}
    \centering
    \includegraphics[width=.85\columnwidth]{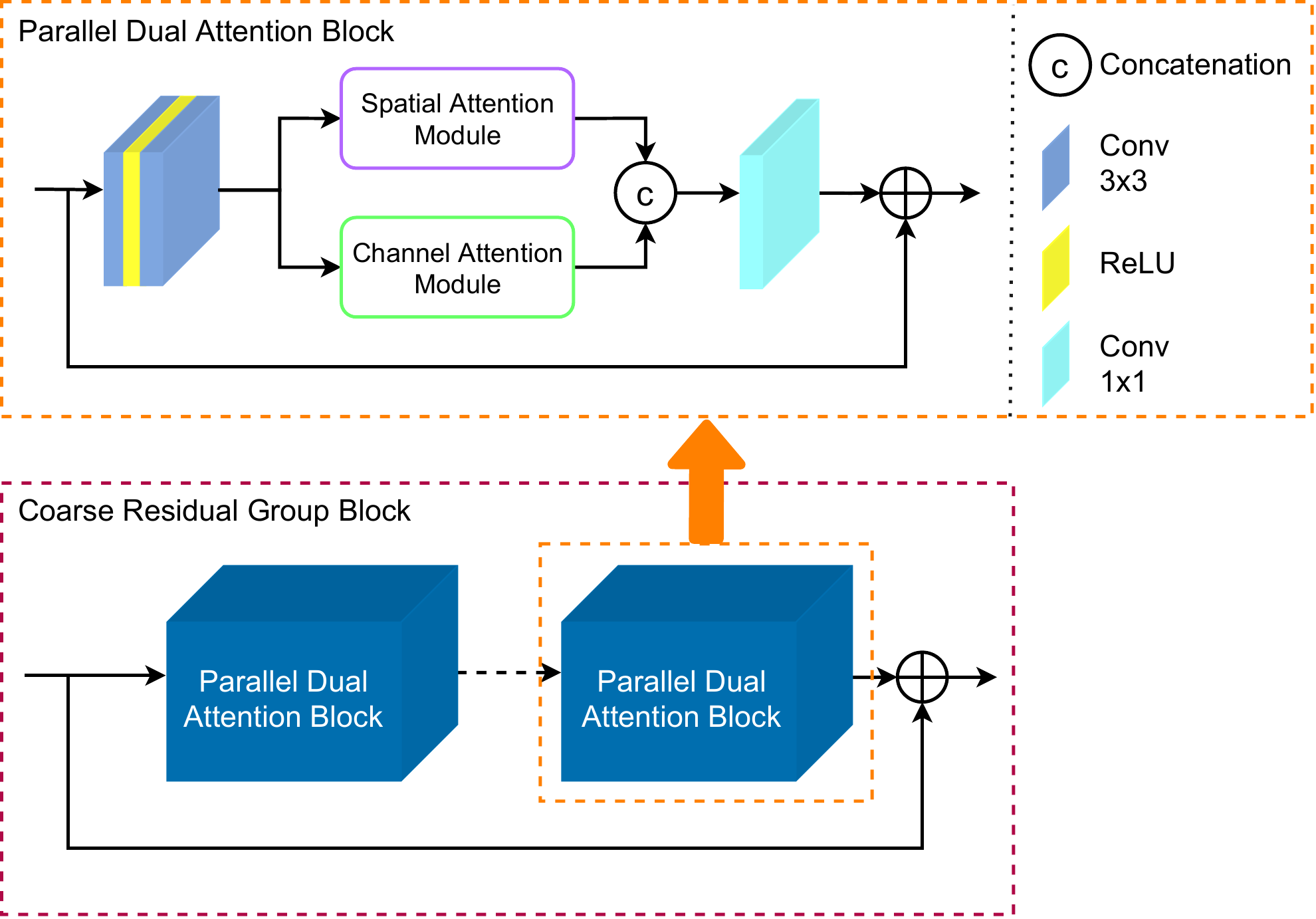}
    \caption{Coarse Residual Group consisting of several Parallel Dual Attention Blocks.}
    \label{fig:coarse_rg}
\end{figure}

By following the channel attention module from \cite{zhang2018image}, we derive two variations of dual attention block by using channel and spatial attention. 
The first variation is a parallel dual attention block. 
This block is used for Residual Group (RG) in the coarse network. 
The detailed architecture of this RG and the parallel dual attention block can be seen in Figure \ref{fig:coarse_rg}.

Studies conducted on \cite{woo2018cbam} found that parallel design cannot outperform sequential design. 
However, in the case of low-level computer vision, giving the same weight to spatial attention and channel attention have their own benefits. 
For example, in the case of SISR for compressed images, there will be many blocky artifacts with the wrong texture. 
This means the spatial attention module will have great benefits to help the network to look where the important and correct texture or color are and ignore the location where it contains incorrect texture information. 
The channel attention module also can help greatly by telling the network 'what' to look at the compressed image. 
For example, it can ignore incorrect texture information or color information and strengthen important texture information. 
If we choose a sequential design, whether the first module is spatial or channel, then sometimes it can contain incorrect information on 'where' to attend or 'what' to attend. 
This makes the sequential design more error-prone compared to the parallel design. 
Because it can contain error, we also employ an additional $1 \times 1$ convolutional layer to help the network choose the correct information on 'what' to attend and 'where' to look, and strengthen the correct information and dampen the incorrect information. 
Hence, we employ a parallel dual attention block for the coarse super-resolution network to significantly reduce many artifacts and retain the most important information from compressed images. 
We also found out that this derivation is similar to the RRG block that CycleISP \cite{zamir2020cycleisp} used.

\begin{figure}
    \centering
    \includegraphics[width=.95\columnwidth]{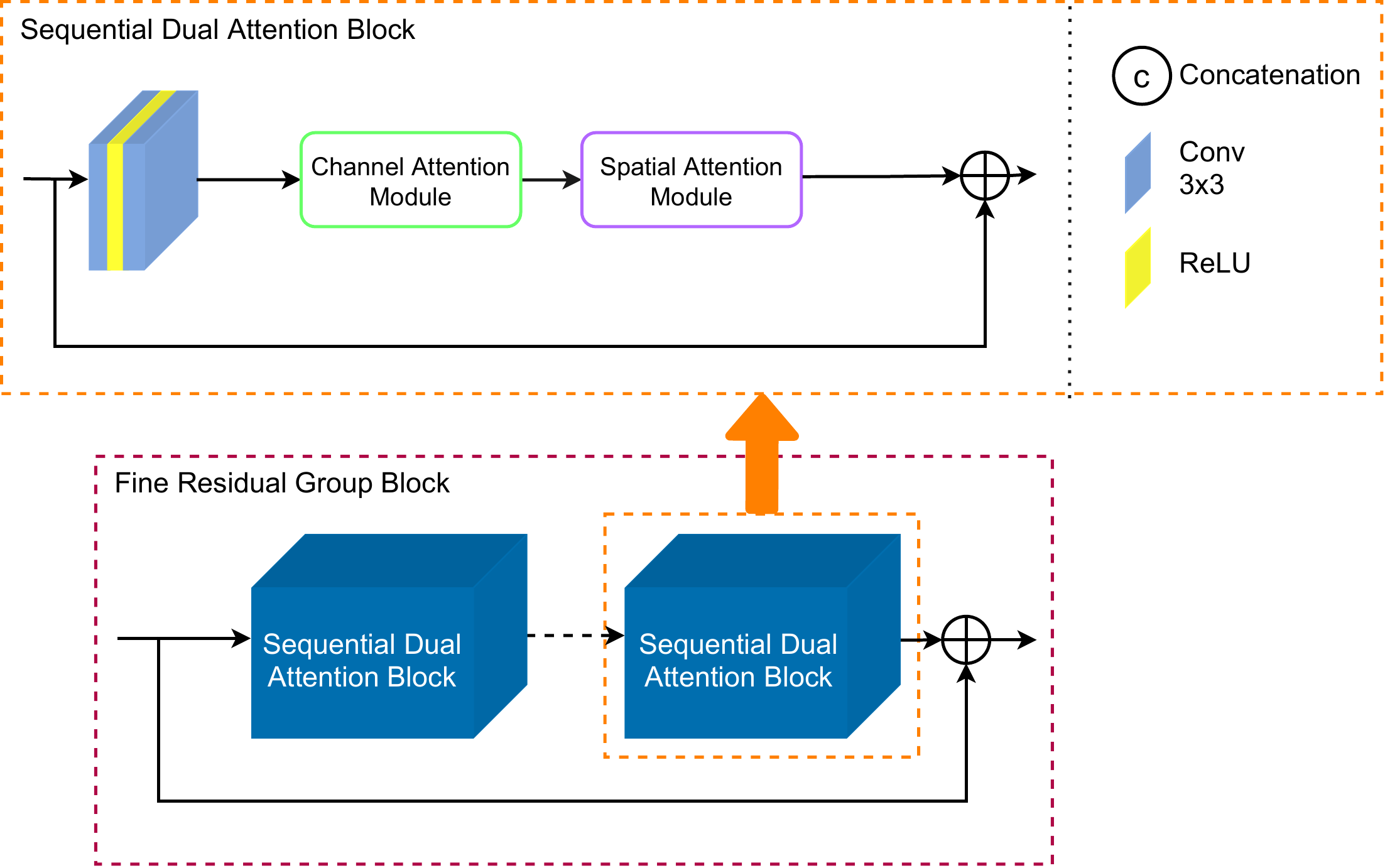}
    \caption{Fine Residual Group consisting of several Sequential Dual Attention Blocks.}
    \label{fig:fine_rg}
\end{figure}

The second derivation is sequential dual attention block. 
This block is used for the RG in the refinement network. 
The detailed architecture of the RG and sequential dual attention block can be seen in Figure \ref{fig:fine_rg}. 
The refinement network input is the output of a coarse super-resolution network. 
This input contains less noise and artifacts compared to the compressed image. 
Because of this, we choose a sequential design of channel attention and spatial attention. 
This arrangement will weigh on 'what' to attend more to make the network focus on reconstructing high-frequency details and texture. 
Even though the spatial information becomes less important, employing spatial attention after channel attention also can help the network to look 'where' is the information of high-frequency details that can be used by the network to help for the reconstruction.

\subsection{Implementation Details}
\label{subsec:implementation_details}
CISRNet employs a two-stage coarse-to-fine learning framework. 
First, the coarse super-resolution network will be trained in the first stage. 
After that, in the second stage, the coarse super-resolution network and refinement network will train jointly in one epoch.
For the first stage, the network is trained for 100 epochs with an initial learning rate of $10^{-4}$ and decreased by a factor of 10 after 75 epochs.
As previously mentioned, the loss function used in the first stage training is L1 weighted by 0.1.
Meanwhile, for the second stage, the network is trained for 200 epochs with an initial learning rate of $0.75 \times 10^{-4}$ and decreased by a  factor of 10 after 150 epochs. 
The second stage training uses content loss where $\lambda_{L1}$ and $\lambda_{p}$ are set as 1 and 0.05 respectively.

All convolutional layers have the output feature of 96. 
For the coarse super-resolution network, 4 Coarse RG is used. 
Each Coarse RG uses 8 parallel dual attention blocks. 
Meanwhile, the refinement network uses 2 Fine RG, where each residual group uses 8 sequential dual attention blocks. 
The reduction factor of the channel attention is 8.

The mini-batch size used for training is 16 with $48 \times 48$ random LR patches for each input.
Each input will be normalized to [0, 1].
The data will be augmented using horizontal and vertical flip, and $90^{\circ}$ rotation.
The training uses Adam \cite{kingma2014adam} as the optimizer.

\section{Experiment}
\label{sec:experiment}

\begin{table*}[ht]
    \centering
    \resizebox{\textwidth}{!}{
    \begin{tabular}{|c|c|c|c|c|c|c|c|}
    \hline
        \textbf{Dataset} & 
        \textbf{Scale} & 
        \textbf{Set5} &
        \begin{tabular}{@{}c@{}}\textbf{Set14} \\ \end{tabular} &
        \begin{tabular}{@{}c@{}}\textbf{B100} \\ \end{tabular} &
        \begin{tabular}{@{}c@{}}\textbf{Urban100} \\ \end{tabular} & 
        \begin{tabular}{@{}c@{}}\textbf{Manga109} \\ \end{tabular} &
        \begin{tabular}{@{}c@{}}\textbf{Complexity} \\ Params (mil.) / Time(s)\end{tabular} \\
    \hline
    \hline
        \textbf{VDSR} \cite{kim2016accurate} &
        \begin{tabular}{@{}c@{}}x2 \\ x3 \\ x4\end{tabular} & 
        \begin{tabular}{@{}c@{}}\textbf{29.138}/0.807 \\ 24.925/0.668 \\ \textbf{26.564}/\textbf{0.725}\end{tabular} & 
        \begin{tabular}{@{}c@{}}25.514/\underline{0.660} \\ 23.542/0.576 \\ 22.919/0.544\end{tabular} &
        \begin{tabular}{@{}c@{}}24.983/0.617 \\ 24.054/0.561 \\ 23.263/0.528\end{tabular} & 
        \begin{tabular}{@{}c@{}}23.194/0.642 \\ 21.541/0.550 \\ 20.777/0.506\end{tabular} &
        \begin{tabular}{@{}c@{}}25.572/0.771 \\ 23.192/0.689 \\ 21.802/0.642\end{tabular} &
        \begin{tabular}{@{}c@{}}\textbf{0.668}/\textbf{33.91} \\ \textbf{0.668}/\textbf{34.52} \\ \textbf{0.668}/\textbf{34.08}\end{tabular} \\
    \hline
        \textbf{RDN} \cite{zhang2018residual} &
        \begin{tabular}{@{}c@{}}x2 \\ x3 \\ x4\end{tabular} & 
        \begin{tabular}{@{}c@{}}28.817/\underline{0.814} \\ \underline{26.486}/\underline{0.748} \\ 24.987/0.698\end{tabular} & 
        \begin{tabular}{@{}c@{}}26.723/\textbf{0.708} \\ \underline{24.965}/\underline{0.635} \\ \underline{23.827}/\underline{0.594}\end{tabular} &
        \begin{tabular}{@{}c@{}}26.050/\underline{0.659} \\ \underline{24.686}/\underline{0.593} \\ \underline{23.812}/\textbf{0.558}\end{tabular} & 
        \begin{tabular}{@{}c@{}}24.775/\underline{0.724} \\ \underline{22.861}/\underline{0.633} \\ \underline{21.744}/\underline{0.577}\end{tabular} &
        \begin{tabular}{@{}c@{}}\underline{25.988}/\underline{0.835} \\ \textbf{22.627}/\underline{0.744} \\ \textbf{20.414}/\textbf{0.672}\end{tabular} &
        \begin{tabular}{@{}c@{}}22.123/\underline{207.36} \\ 22.308/\underline{95.96} \\ 22.271/\underline{56.72}\end{tabular} \\
    \hline
        \textbf{RCAN} \cite{zhang2018image} & 
        \begin{tabular}{@{}c@{}}x2 \\ x3 \\ x4\end{tabular} & 
        \begin{tabular}{@{}c@{}}28.929/\textbf{0.816} \\ -/- \\ 24.191/0.640\end{tabular} & 
        \begin{tabular}{@{}c@{}}\textbf{26.784}/\textbf{0.708} \\ -/- \\ 23.215/0.561\end{tabular} &
        \begin{tabular}{@{}c@{}}\textbf{26.098}/\textbf{0.660} \\ -/- \\ 23.466/\underline{0.537}\end{tabular} & 
        \begin{tabular}{@{}c@{}}\textbf{24.941}/\textbf{0.727} \\ -/- \\ 21.088/0.526\end{tabular} &
        \begin{tabular}{@{}c@{}}25.859/0.833 \\ -/- \\ 20.657/0.628\end{tabular} &
        \begin{tabular}{@{}c@{}}15.444/284.76 \\ 15.629/187.13 \\ 15.592/202.15\end{tabular} \\
    \hline
        \textbf{CISRNet} (ours) &
        \begin{tabular}{@{}c@{}}x2 \\ x3 \\ x4\end{tabular} & 
        \begin{tabular}{@{}c@{}}\underline{28.936}/\textbf{0.816} \\ \textbf{26.579}/\textbf{0.753} \\ \underline{25.026}/\underline{0.702}\end{tabular} & 
        \begin{tabular}{@{}c@{}}\underline{26.777}/\textbf{0.708} \\ \textbf{25.001}/\textbf{0.636} \\ \textbf{23.880}/\textbf{0.596}\end{tabular} &
        \begin{tabular}{@{}c@{}}\underline{26.081}/\underline{0.659} \\ \textbf{24.714}/\textbf{0.594} \\ \textbf{23.827}/\textbf{0.558}\end{tabular} & 
        \begin{tabular}{@{}c@{}}\underline{24.932}/\textbf{0.727} \\ \textbf{23.004}/\textbf{0.638} \\ \textbf{21.861}/\textbf{0.582}\end{tabular} &
        \begin{tabular}{@{}c@{}}\textbf{26.013}/\textbf{0.836} \\ \underline{22.585}/\textbf{0.747} \\ \underline{20.210}/\underline{0.671}\end{tabular} &
        \begin{tabular}{@{}c@{}}\underline{9.6}/327.34 \\ \underline{10.015}/179.60 \\ \underline{9.932}/163.75\end{tabular} \\
    \hline
    \end{tabular}
    }
    \caption{Quantitative result (average PSNR/SSIM) for compressed image super resolution on various datasets. \textbf{Bold} and \underline{underline} denotes the best and second-best performance respectively.}
    \label{table:main_quantitative_result}
\end{table*}

\noindent \textbf{Dataset and baseline.} The training dataset is 800 training images from DIV2K dataset \cite{agustsson2017ntire}. For testing, Set5 \cite{bevilacqua2012low}, Set14 \cite{zeyde2010single}, B100 \cite{martin2001database}, Urban100 \cite{huang2015single}, and Manga109 \cite{matsui2017sketch} are used. All of the low-resolution images in the training set and the test set are compressed with a quality of $10\%$ from the original image. For the baseline, three state-of-the-art models for single-image super resolution such as VDSR \cite{kim2016accurate}, RDN \cite{zhang2018residual} and RCAN \cite{zhang2018image} are retrained using the compressed dataset with their respective training and implementation detail. 
There is no two-stage network used for the baseline because we found that the result is below the existing baseline. 
The same condition applies for CISRDCNN \cite{chen2018cisrdcnn}, where the result is below the chosen baseline for the quality factor of $10\%$. 
We perform our experiments for three upscaling factors; 2, 3, and 4, each denoted as x2, x3, and x4 in this paper.

\subsection{Comparison with Existing SISR Method}
\label{sec:comparison_with_baselines}

\begin{figure*}[t]
\centering
\includegraphics[width=\textwidth]{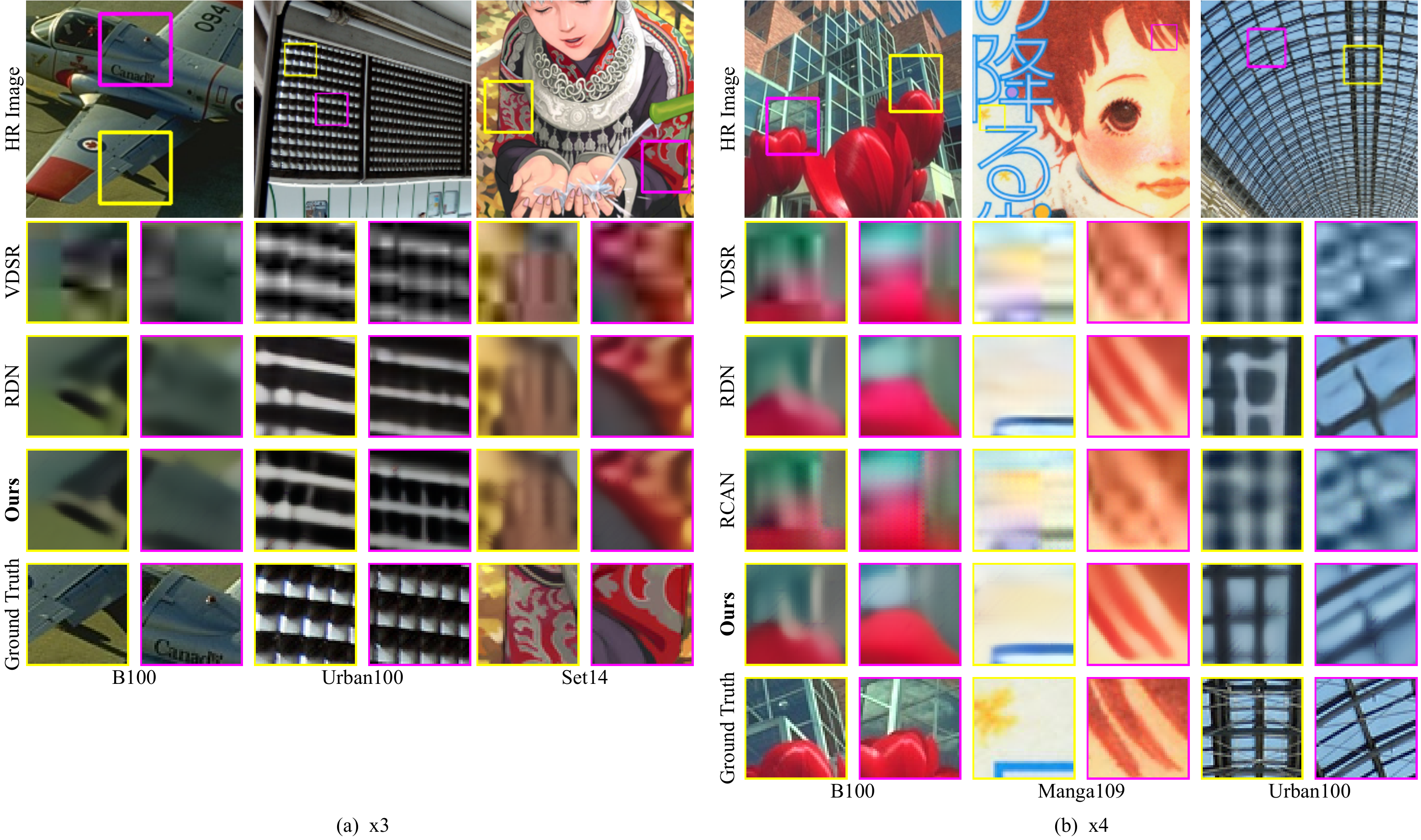}
   \caption{Qualitative result of our method compared to others.}
\label{fig:qualitative_result}
\end{figure*}

\noindent \textbf{Quantitative comparison.} 
We compare the Y-channel PSNR in dB and SSIM for various methods on all 328 images from the aforementioned testing datasets.
We also incorporate the number of parameters in millions and inference time in seconds for all testing datasets to measure model complexity.
An NVIDIA Titan Xp is utilized for performing inference on each method.
Our method achieves the second-best in terms of the number of parameters, whereas our inference time is relatively higher compared to other methods with a higher number of parameters. 
This is because our refinement network utilizes a full high-resolution image to output the final result.
For RCAN \cite{zhang2018image}, we leave our results blank in x3 due to technical issues from the official repository during inference.

In terms of PSNR and SSIM, our method outperforms other methods on scale factor of 3 and 4 in Set14 \cite{zeyde2010single}, B100 \cite{martin2001database}, and Urban100 \cite{huang2015single} datasets. 
This set of experiments and datasets have a higher resemblance to scenarios commonly seen in a real-world setting compared to the others.
Better PSNR and SSIM mean our method is highly capable of reducing the noise and also recovering high-frequency details in more complex environments.
Our method improves around 0.01 to 0.1 dB in terms of PSNR compared to the second-best method, with the largest gain coming from the scenario of x4 in Urban100 \cite{huang2015single} dataset, which is arguably the most complex set of tasks in these experiments. 
Also, note that in compressed image super-resolution, there exists many blocky artifacts and noise subjected to removal within an image that may contain cues to recover its texture and detail.
This further implies that methods with high performance are able to successfully remove the noise and can differentiate between the noise and the cue that can be used to preserve some complex texture and high-frequency details.

Even though our method has worse performance compared to other methods on x2 experiments on most datasets, we believe that this setting is not an appropriate way to measure the performance of compressed image super-resolution methods given the lack of complexity.
VDSR \cite{kim2016accurate} achieves the best performance on Set5 \cite{bevilacqua2012low} for x2 and x4 scenario, but then achieves significantly lower performances on other experiments with more complex datasets compared to the other methods.
We also want to point out that Manga109 \cite{matsui2017sketch} dataset contains a set of manga covers that offers less complexity in terms of texture and detail compared to other datasets such as B100 \cite{martin2001database} and Urban100 \cite{huang2015single}.
From all of these results, we can conclude that our method achieves dominant performance in environments resembling real-world scenarios and competitive performance in less complex environments.

\noindent \textbf{Qualitative comparison.}
Figure \ref{fig:qualitative_result} visualizes the result of our method compared to other methods.
Here, we may observe that our method is capable of reconstructing high-frequency details and sharp edges. 
In addition, our method does not create excessive blurry artifacts compared to other methods. 
This result further supports our claim in the previous point where our method can successfully remove the noise and preserve some complex texture and high-frequency details, and perform relatively better than other baselines. 
Despite this, the problem of compressed image super-resolution still remains challenging. 
Figure \ref{fig:qualitative_result} also suggests that all methods in this experiment still suffer from the inability to produce correct high-frequency details, texture, and structure from input with many blocky artifacts.

\subsection{Result on Real-World Compressed Image}
\begin{figure}[ht]
\centering
\includegraphics[width=\columnwidth]{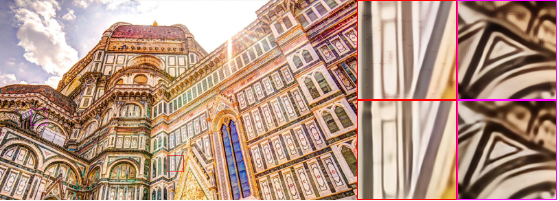}
(a) The input (left), our result (top), and RDN result (bottom) with scale factor of 3. \cite{shopping2018cathedral}
\includegraphics[width=\columnwidth]{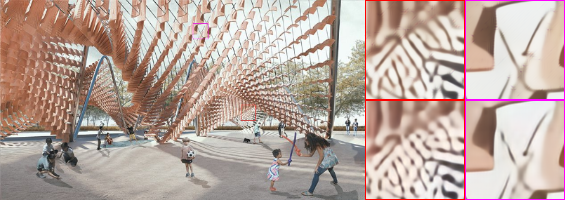}
(b) The input (left), our result (top), and RDN result (bottom) with scale factor of 3. \cite{domusweb2018flocking}
   \caption{The qualitative result on real images.}
\label{fig:real_images_result}
\end{figure}

To further evaluate the performance of our model with other baselines, we perform additional inference with real-world compressed images. 
We collect two real-world images containing complex structures from the internet. 
After that, we perform compression on these two images by decreasing the image quality to 10\% of the original. 
These images will become the input to our method and RDN as the competitive baseline. 
The result between our method and RDN can be seen in Figure \ref{fig:real_images_result}. 
As it can be seen in Figure \ref{fig:real_images_result} (a), RDN cannot reconstruct the complex structure and appears to be more blurry (red patch), while our method can retain the complex structure and does not remove many high-frequency details. 
This explanation also applies for the result in Figure \ref{fig:real_images_result} (b).

\subsection{Ablation Study}
In this section, we analyze the contribution of different components within our network. 
It will mainly focus on the number of residual groups, the dual attention block, the loss function, and the contribution of the refinement network.

\noindent \textbf{Effect of different number of residual group for coarse \& refinement network architecture.} 
We study the effect of different numbers of residual groups (RG) for coarse network and refinement network. 
In this experiment, we use the same learning strategy and hyperparameter described in Section \ref{subsec:implementation_details}.
The first row in Table \ref{table:number_of_blocks} indicates the performance from the coarse network (without refinement) with a different number of RGs.
Meanwhile, the second row in Table \ref{table:number_of_blocks} denotes the performance from the full network consisting of coarse network with four RGs and refinement network with a varying number of RGs. 

\begin{table}[hb]
    \centering
    \begin{tabular}{c c c}
    \toprule
    \textbf{Network} & \textbf{Number of RG} &  \textbf{PSNR / SSIM}\\
    \midrule
        Coarse & 
        \begin{tabular}{@{}c@{}}3 \\ 4 \\ 5 \\ 6\end{tabular} & \begin{tabular}{@{}c@{}}26.049/0.658 \\ 26.054/0.658 \\ 26.059/0.658 \\ \textbf{26.067}/\textbf{0.659}\end{tabular} \\
    \midrule
        Refinement & 
        \begin{tabular}{@{}c@{}}1 \\ 2 \\ 3\end{tabular} &
        \begin{tabular}{@{}c@{}}26.079/\textbf{0.659} \\ 26.081/\textbf{0.659} \\ \textbf{26.082}/\textbf{0.659}\end{tabular}\\
    \bottomrule
    \end{tabular}
    \caption{Result of using a different number of residual group (RG) for coarse network and refinement network with a scale factor of x2 on B100  dataset.}
    \label{table:number_of_blocks}
\end{table}

According to the first row in Table \ref{table:number_of_blocks}, by adding the number of residual groups, the performance can be improved.
The result increased by 0.005 dB in terms of PSNR by adding a residual group from 3 to 4.
However, when the number of residual groups is increased further, the result increased again by around 0.008 dB.
We don't do further experiments because we also need to take into account the complexity of our method.
Therefore, we choose 4 as the number of residual groups for the coarse network which achieves the equilibrium with respect to performance and complexity.

In terms of refinement network, adding the number of residual groups results in a slight performance improvement.
As it can be noticed in the second row of Table \ref{table:number_of_blocks}, by increasing the number of residual groups from one to two, the performance only increases by 0.002 dB, which is not that substantial.
Furthermore, using three RGs in the refinement network increases performance by only 0.001 dB.
Since we believe that increasing the number of RGs further doesn't increase performance by a significant amount, we only used two RGs for our final network.

\noindent \textbf{Different dual attention block for each network.}
In this section, we further verify our design choice of using the Parallel Dual Attention Block (PDAB) for the coarse network and the Sequential Dual Attention Block (SDAB) for the refinement network.
The main reason for using PDAB for the refinement network and SDAB for the coarse network can be seen in Section \ref{sec:residual_group}.
The first row of Table \ref{table:different_dual_attention_block} represents the result of the coarse network according to the setup in Section \ref{subsec:implementation_details} under two different block types.
Meanwhile, the second row of Table \ref{table:different_dual_attention_block} represents the result of the full network under the configuration specified in Section \ref{subsec:implementation_details} with two different blocks within the refinement network. These results denote the performance of respective networks on B100 \cite{martin2001database} for x2 scenario.

\begin{table}[b]
    \centering
    \begin{tabular}{c c c}
    \toprule
    \textbf{Network} & \textbf{Type of Block} &  \textbf{PSNR / SSIM}\\
    \midrule
        Coarse & 
        \begin{tabular}{@{}c@{}}PDAB \\ SDAB\end{tabular} & \begin{tabular}{@{}c@{}}\textbf{26.054}/\textbf{0.658} \\ 26.034/0.657\end{tabular} \\
    \midrule
        Refinement & 
        \begin{tabular}{@{}c@{}}PDAB \\ SDAB\end{tabular} & \begin{tabular}{@{}c@{}}26.056/0.658 \\ \textbf{26.081}/\textbf{0.659}\end{tabular}\\ 
    \bottomrule
    \end{tabular}
    \caption{Result of using different type of dual attention block for coarse super-resolution and refinement network with a scale factor of x2 on B100  dataset.}
    \label{table:different_dual_attention_block}
\end{table}
According to Table \ref{table:different_dual_attention_block}, the combination of PDAB for coarse network and SDAB for the refinement network can achieve the best performance.
The performance for using different blocks can decrease the performance in terms of PSNR by 0.02 dB for the coarse network and 0.025 dB for the refinement network.
We also find that using PDAB for the refinement network leads to gradient explosion.
We believe that the refinement network should not have complex architecture.
This is further supported by our claim in the first ablation study, where the performance increase is decreasing as the number of residual groups increases.

\begin{figure}[t]
\centering
\includegraphics[width=\columnwidth]{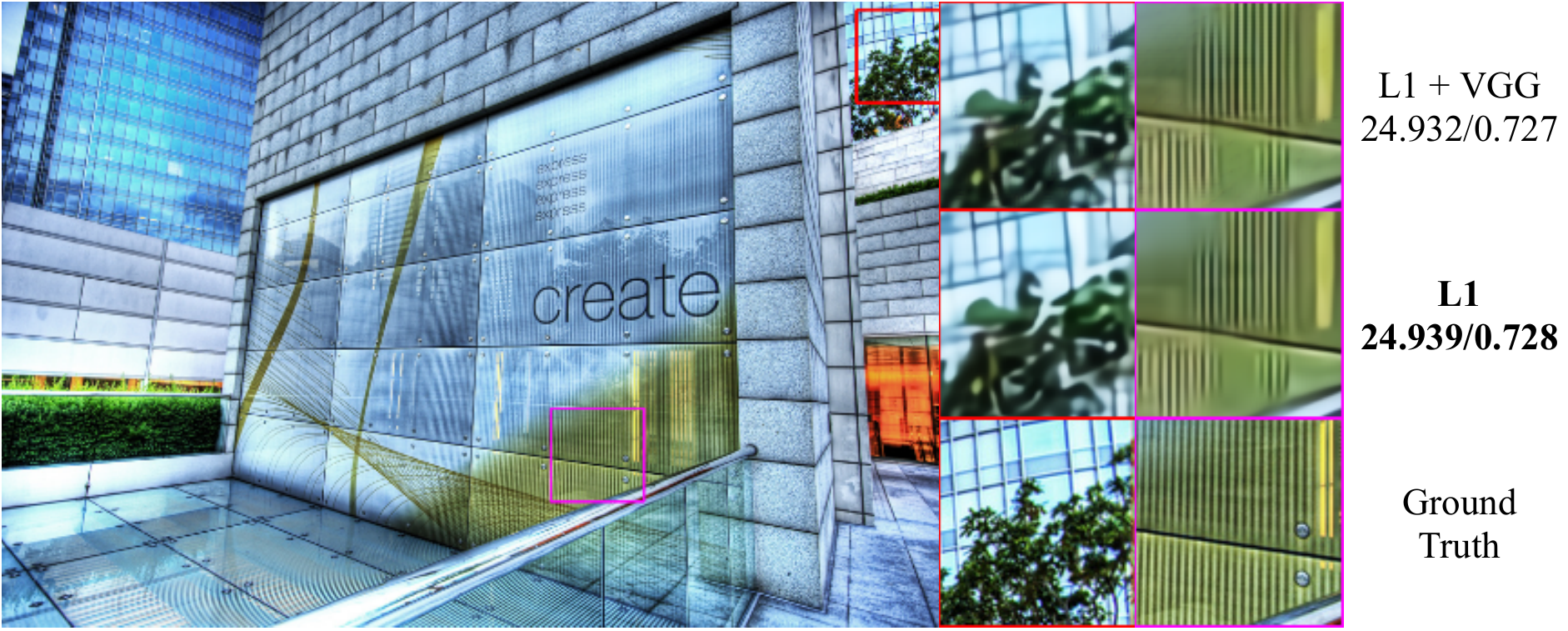}
   \caption{Quantitative and qualitative results using different loss functions. The number below the loss function denotes PSNR/SSIM in Urban100 dataset for x2. The left image is the ground truth high-resolution image.}
\label{fig:loss_function_ablation}
\end{figure}

\noindent \textbf{Loss function.}
In this study, we compare the result of using VGG and L1 as the loss function and only L1 as the loss function.
Figure \ref{fig:loss_function_ablation} shows the result of using different loss functions, where using the L1 loss function achieves better performance both in terms of PSNR and SSIM.
However, we found that qualitatively, applying VGG loss enhance the ability of the network to reconstruct high-frequency details and complex structure.
A similar result also can be seen in \cite{wang2018esrgan}, where VGG loss decreases the performance in terms of PSNR and SSIM but increases the ability of the network to reconstruct high-frequency details and texture.

\begin{figure}[t]
\centering
\includegraphics[width=\columnwidth]{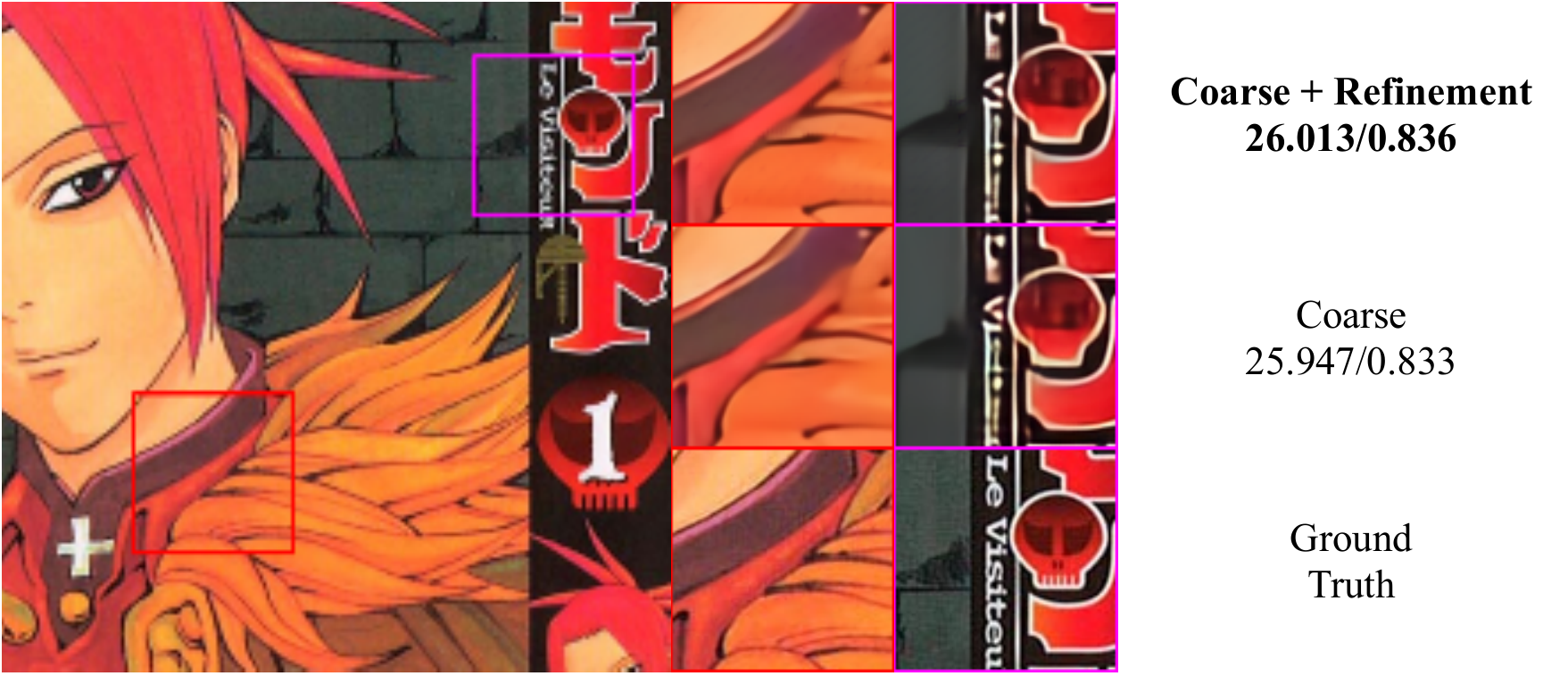}
   \caption{Quantitative and qualitative results using only coarse network and coarse plus refinement network. The number below the network denotes PSNR/SSIM in Manga109 dataset for x2. The left image is the ground truth high-resolution image.}
\label{fig:refinement_ablation}
\end{figure}
\noindent \textbf{Refinement network contribution.}
The quantitative and qualitative result of using coarse network only for the full network and coarse with refinement network as the full network can be seen in Figure \ref{fig:refinement_ablation}.
By using a refinement network, quantitatively it can improve the performance by 0.066 dB and 0.003 in terms of PSNR and SSIM.
Qualitatively, by using a refinement network, the resulting image has a sharper edge and correct high-frequency details (the dark region in the collar). 
Meanwhile, by using only a coarse network, the resulting image has distorted texture and incorrect high-frequency details.
This result is in accordance with our goal that incorporates a refinement network to sharpen the image, reconstruct the texture and high-frequency detail correctly.

\section{Conclusion and Future Works}
In this paper, we present CISRNet to tackle the problem of compressed image super-resolution.
This method employs a two-stage coarse-to-fine learning framework. 
Specifically, we employ recursive learning schema for the coarse super-resolution network and residual learning for the refinement network.
In addition, we also propose two different types of dual attention blocks to be used in our network.
We also show that our network achieves competitive performance in the less complex dataset and outperforms other methods in real-scene and more complex datasets.
Even though we can achieve better performance, we also find that this problem still remains challenging.
Our method and other baseline methods still cannot reconstruct the texture and structure of the super-resolution version of the compressed image correctly.
This opens a new possibility where this problem needs to be tackled by using a generative method (e.g. GAN) instead of a conventional network with supervised learning.
The code and additional qualitative results can be seen in our Github repository\footnote{\href{https://github.com/agusgun/cisr}{https://github.com/agusgun/cisr}}.

{\small
\bibliographystyle{ieee_fullname}
\bibliography{reference}

\begin{thebibliography}{10}\itemsep=-1pt

\bibitem{agustsson2017ntire}
Eirikur Agustsson and Radu Timofte.
\newblock Ntire 2017 challenge on single image super-resolution: Dataset and
  study.
\newblock In {\em Proceedings of the IEEE Conference on Computer Vision and
  Pattern Recognition Workshops}, pages 126--135, 2017.

\bibitem{bashir2021comprehensive}
Syed Muhammad~Arsalan Bashir, Yi Wang, and Mahrukh Khan.
\newblock A comprehensive review of deep learning-based single image
  super-resolution.
\newblock {\em arXiv preprint arXiv:2102.09351}, 2021.

\bibitem{bates2007multicolor}
Mark Bates, Bo Huang, Graham~T Dempsey, and Xiaowei Zhuang.
\newblock Multicolor super-resolution imaging with photo-switchable fluorescent
  probes.
\newblock {\em Science}, 317(5845):1749--1753, 2007.

\bibitem{bevilacqua2012low}
Marco Bevilacqua, Aline Roumy, Christine Guillemot, and Marie~Line
  Alberi-Morel.
\newblock Low-complexity single-image super-resolution based on nonnegative
  neighbor embedding.
\newblock 2012.

\bibitem{chen2018cisrdcnn}
Honggang Chen, Xiaohai He, Chao Ren, Linbo Qing, and Qizhi Teng.
\newblock Cisrdcnn: Super-resolution of compressed images using deep
  convolutional neural networks.
\newblock {\em Neurocomputing}, 285:204--219, 2018.

\bibitem{cheng2015deep}
Zezhou Cheng, Qingxiong Yang, and Bin Sheng.
\newblock Deep colorization.
\newblock In {\em Proceedings of the IEEE International Conference on Computer
  Vision}, pages 415--423, 2015.

\bibitem{domusweb2018flocking}
{Domusweb}.
\newblock Flocking tejas pavilion, 2018.
\newblock [Online; accessed May 22, 2021].

\bibitem{dong2015compression}
Chao Dong, Yubin Deng, Chen~Change Loy, and Xiaoou Tang.
\newblock Compression artifacts reduction by a deep convolutional network.
\newblock In {\em Proceedings of the IEEE International Conference on Computer
  Vision}, pages 576--584, 2015.

\bibitem{dong2014learning}
Chao Dong, Chen~Change Loy, Kaiming He, and Xiaoou Tang.
\newblock Learning a deep convolutional network for image super-resolution.
\newblock In {\em European conference on computer vision}, pages 184--199.
  Springer, 2014.

\bibitem{dong2015image}
Chao Dong, Chen~Change Loy, Kaiming He, and Xiaoou Tang.
\newblock Image super-resolution using deep convolutional networks.
\newblock {\em IEEE transactions on pattern analysis and machine intelligence},
  38(2):295--307, 2015.

\bibitem{foi2007pointwise}
Alessandro Foi, Vladimir Katkovnik, and Karen Egiazarian.
\newblock Pointwise shape-adaptive dct for high-quality denoising and
  deblocking of grayscale and color images.
\newblock {\em IEEE transactions on image processing}, 16(5):1395--1411, 2007.

\bibitem{hu2016efficient}
Yinlin Hu, Rui Song, and Yunsong Li.
\newblock Efficient coarse-to-fine patchmatch for large displacement optical
  flow.
\newblock In {\em Proceedings of the IEEE Conference on Computer Vision and
  Pattern Recognition}, pages 5704--5712, 2016.

\bibitem{huang2015single}
Jia-Bin Huang, Abhishek Singh, and Narendra Ahuja.
\newblock Single image super-resolution from transformed self-exemplars.
\newblock In {\em Proceedings of the IEEE conference on computer vision and
  pattern recognition}, pages 5197--5206, 2015.

\bibitem{khosla2020supervised}
Prannay Khosla, Piotr Teterwak, Chen Wang, Aaron Sarna, Yonglong Tian, Phillip
  Isola, Aaron Maschinot, Ce Liu, and Dilip Krishnan.
\newblock Supervised contrastive learning.
\newblock {\em arXiv preprint arXiv:2004.11362}, 2020.

\bibitem{kim2016accurate}
Jiwon Kim, Jung~Kwon Lee, and Kyoung~Mu Lee.
\newblock Accurate image super-resolution using very deep convolutional
  networks.
\newblock In {\em Proceedings of the IEEE conference on computer vision and
  pattern recognition}, pages 1646--1654, 2016.

\bibitem{kim2016deeply}
Jiwon Kim, Jung~Kwon Lee, and Kyoung~Mu Lee.
\newblock Deeply-recursive convolutional network for image super-resolution.
\newblock In {\em Proceedings of the IEEE conference on computer vision and
  pattern recognition}, pages 1637--1645, 2016.

\bibitem{kingma2014adam}
Diederik~P Kingma and Jimmy Ba.
\newblock Adam: A method for stochastic optimization.
\newblock {\em arXiv preprint arXiv:1412.6980}, 2014.

\bibitem{ledig2017photo}
Christian Ledig, Lucas Theis, Ferenc Husz{\'a}r, Jose Caballero, Andrew
  Cunningham, Alejandro Acosta, Andrew Aitken, Alykhan Tejani, Johannes Totz,
  Zehan Wang, et~al.
\newblock Photo-realistic single image super-resolution using a generative
  adversarial network.
\newblock In {\em Proceedings of the IEEE conference on computer vision and
  pattern recognition}, pages 4681--4690, 2017.

\bibitem{liu2020comprehensive}
Jiaying Liu, Dong Liu, Wenhan Yang, Sifeng Xia, Xiaoshuai Zhang, and Yuanying
  Dai.
\newblock A comprehensive benchmark for single image compression artifact
  reduction.
\newblock {\em IEEE Transactions on Image Processing}, 29:7845--7860, 2020.

\bibitem{lucas1981iterative}
Bruce~D Lucas, Takeo Kanade, et~al.
\newblock An iterative image registration technique with an application to
  stereo vision.
\newblock Vancouver, British Columbia, 1981.

\bibitem{ma2019coarse}
Yuqing Ma, Xianglong Liu, Shihao Bai, Lei Wang, Dailan He, and Aishan Liu.
\newblock Coarse-to-fine image inpainting via region-wise convolutions and
  non-local correlation.
\newblock In {\em IJCAI}, pages 3123--3129, 2019.

\bibitem{martin2001database}
David Martin, Charless Fowlkes, Doron Tal, and Jitendra Malik.
\newblock A database of human segmented natural images and its application to
  evaluating segmentation algorithms and measuring ecological statistics.
\newblock In {\em Proceedings Eighth IEEE International Conference on Computer
  Vision. ICCV 2001}, volume~2, pages 416--423. IEEE, 2001.

\bibitem{matsui2017sketch}
Yusuke Matsui, Kota Ito, Yuji Aramaki, Azuma Fujimoto, Toru Ogawa, Toshihiko
  Yamasaki, and Kiyoharu Aizawa.
\newblock Sketch-based manga retrieval using manga109 dataset.
\newblock {\em Multimedia Tools and Applications}, 76(20):21811--21838, 2017.

\bibitem{ono2013optimized}
Shunsuke Ono and Isao Yamada.
\newblock Optimized jpeg image decompression with super-resolution
  interpolation using multi-order total variation.
\newblock In {\em 2013 IEEE International Conference on Image Processing},
  pages 474--478. IEEE, 2013.

\bibitem{o2019deep}
Niall O’Mahony, Sean Campbell, Anderson Carvalho, Suman Harapanahalli,
  Gustavo~Velasco Hernandez, Lenka Krpalkova, Daniel Riordan, and Joseph Walsh.
\newblock Deep learning vs. traditional computer vision.
\newblock In {\em Science and Information Conference}, pages 128--144.
  Springer, 2019.

\bibitem{shi2016real}
Wenzhe Shi, Jose Caballero, Ferenc Husz{\'a}r, Johannes Totz, Andrew~P Aitken,
  Rob Bishop, Daniel Rueckert, and Zehan Wang.
\newblock Real-time single image and video super-resolution using an efficient
  sub-pixel convolutional neural network.
\newblock In {\em Proceedings of the IEEE conference on computer vision and
  pattern recognition}, pages 1874--1883, 2016.

\bibitem{shopping2018cathedral}
{Shopping \& Charity}.
\newblock Florence cathedral, 2018.
\newblock [Online; accessed May 22, 2021].

\bibitem{simonyan2014very}
Karen Simonyan and Andrew Zisserman.
\newblock Very deep convolutional networks for large-scale image recognition.
\newblock {\em arXiv preprint arXiv:1409.1556}, 2014.

\bibitem{svoboda2016compression}
Pavel Svoboda, Michal Hradis, David Barina, and Pavel Zemcik.
\newblock Compression artifacts removal using convolutional neural networks.
\newblock {\em arXiv preprint arXiv:1605.00366}, 2016.

\bibitem{tian2020coarse}
Chunwei Tian, Yong Xu, Wangmeng Zuo, Bob Zhang, Lunke Fei, and Chia-Wen Lin.
\newblock Coarse-to-fine cnn for image super-resolution.
\newblock {\em IEEE Transactions on Multimedia}, 2020.

\bibitem{wang2018esrgan}
Xintao Wang, Ke Yu, Shixiang Wu, Jinjin Gu, Yihao Liu, Chao Dong, Yu Qiao, and
  Chen Change~Loy.
\newblock Esrgan: Enhanced super-resolution generative adversarial networks.
\newblock In {\em Proceedings of the European Conference on Computer Vision
  (ECCV) Workshops}, pages 0--0, 2018.

\bibitem{wang2020deep}
Zhihao Wang, Jian Chen, and Steven~CH Hoi.
\newblock Deep learning for image super-resolution: A survey.
\newblock {\em IEEE transactions on pattern analysis and machine intelligence},
  2020.

\bibitem{wong2009document}
Tak-Shing Wong, Charles~A Bouman, Ilya Pollak, and Zhigang Fan.
\newblock A document image model and estimation algorithm for optimized jpeg
  decompression.
\newblock {\em IEEE Transactions on Image Processing}, 18(11):2518--2535, 2009.

\bibitem{woo2018cbam}
Sanghyun Woo, Jongchan Park, Joon-Young Lee, and In~So Kweon.
\newblock Cbam: Convolutional block attention module.
\newblock In {\em Proceedings of the European conference on computer vision
  (ECCV)}, pages 3--19, 2018.

\bibitem{wu2019liteeval}
Zuxuan Wu, Caiming Xiong, Yu-Gang Jiang, and Larry~S Davis.
\newblock Liteeval: A coarse-to-fine framework for resource efficient video
  recognition.
\newblock {\em arXiv preprint arXiv:1912.01601}, 2019.

\bibitem{xiao2012single}
Jia Xiao, Chen Wang, and Xiyuan Hu.
\newblock Single image super-resolution in compressed domain based on field of
  expert prior.
\newblock In {\em 2012 5th International Congress on Image and Signal
  Processing}, pages 607--611. IEEE, 2012.

\bibitem{yu2018generative}
Jiahui Yu, Zhe Lin, Jimei Yang, Xiaohui Shen, Xin Lu, and Thomas~S Huang.
\newblock Generative image inpainting with contextual attention.
\newblock In {\em Proceedings of the IEEE conference on computer vision and
  pattern recognition}, pages 5505--5514, 2018.

\bibitem{zamir2020cycleisp}
Syed~Waqas Zamir, Aditya Arora, Salman Khan, Munawar Hayat, Fahad~Shahbaz Khan,
  Ming-Hsuan Yang, and Ling Shao.
\newblock Cycleisp: Real image restoration via improved data synthesis.
\newblock In {\em Proceedings of the IEEE/CVF Conference on Computer Vision and
  Pattern Recognition}, pages 2696--2705, 2020.

\bibitem{zeyde2010single}
Roman Zeyde, Michael Elad, and Matan Protter.
\newblock On single image scale-up using sparse-representations.
\newblock In {\em International conference on curves and surfaces}, pages
  711--730. Springer, 2010.

\bibitem{zhang2017beyond}
Kai Zhang, Wangmeng Zuo, Yunjin Chen, Deyu Meng, and Lei Zhang.
\newblock Beyond a gaussian denoiser: Residual learning of deep cnn for image
  denoising.
\newblock {\em IEEE transactions on image processing}, 26(7):3142--3155, 2017.

\bibitem{zhang2010super}
Liangpei Zhang, Hongyan Zhang, Huanfeng Shen, and Pingxiang Li.
\newblock A super-resolution reconstruction algorithm for surveillance images.
\newblock {\em Signal Processing}, 90(3):848--859, 2010.

\bibitem{zhang2018dmcnn}
Xiaoshuai Zhang, Wenhan Yang, Yueyu Hu, and Jiaying Liu.
\newblock Dmcnn: Dual-domain multi-scale convolutional neural network for
  compression artifacts removal.
\newblock In {\em 2018 25th IEEE International Conference on Image Processing
  (ICIP)}, pages 390--394. IEEE, 2018.

\bibitem{zhang2018image}
Yulun Zhang, Kunpeng Li, Kai Li, Lichen Wang, Bineng Zhong, and Yun Fu.
\newblock Image super-resolution using very deep residual channel attention
  networks.
\newblock In {\em Proceedings of the European conference on computer vision
  (ECCV)}, pages 286--301, 2018.

\bibitem{zhang2018residual}
Yulun Zhang, Yapeng Tian, Yu Kong, Bineng Zhong, and Yun Fu.
\newblock Residual dense network for image super-resolution.
\newblock In {\em CVPR}, 2018.

\end{thebibliography}
}

\end{document}